# Achieving 100,000,000 database inserts per second using Accumulo and D4M

Jeremy Kepner, William Arcand, David Bestor, Bill Bergeron, Chansup Byun, Vijay Gadepally, Matthew Hubbell, Peter Michaleas, Julie Mullen, Andrew Prout, Albert Reuther, Antonio Rosa, Charles Yee

MIT Lincoln Laboratory, Lexington, MA, U.S.A.

*Abstract*—The Apache Accumulo database is an open source relaxed consistency database that is widely used for government applications. Accumulo is designed to deliver high performance on unstructured data such as graphs of network data. This paper tests the performance of Accumulo using data from the Graph500 benchmark. The Dynamic Distributed Dimensional Data Model (D4M) software is used to implement the benchmark on a 216-node cluster running the MIT SuperCloud software stack. A peak performance of over 100,000,000 database inserts per second was achieved which is 100x larger than the highest previously published value for any other database. The performance scales linearly with the number of ingest clients, number of database servers, and data size. The performance was achieved by adapting several supercomputing techniques to this application: distributed arrays, domain decomposition, adaptive load balancing, and single-program-multiple-data programming.

*Keywords-component; Accumulo; Hadoop; Big Data; Graph500; D4M; MIT SuperCloud*

I. INTRODUCTION

Non-traditional, relaxed consistency, triple store databases provide high performance on commodity computing hardware to I/O intensive data mining applications with low data modification requirements. These databases are the backbone of many web companies and they include: Google Big Table [Chang 2008], Amazon Dynamo [DeCandia 2007], Cassandra (cassandra.apache.org), and HBase (hbase.apache.org). The Google Big Table architecture has spawned the development of a wide variety of open source "NoSQL" database implementations [Stonebraker 2010, 2012]. Many of these implementations are built on top of the Hadoop (hadoop.apache.org) distributed computing infrastructure that provides distributed data storage and replication services to these databases. A key element of these databases is relaxed consistency. Traditional databases provide a high level of ACID (atomicity, consistency, isolation, durability). High ACID databases guarantee that separate queries of the same data at the same time will give the same answer. Relaxed consistency databases provide BASE (Basic Availability, Soft-state, Eventual consistency), and guarantee that queries will provide the same answers eventually. In exchange, relaxed consistency databases can be built simply and provide high performance on commodity computing hardware.

The Accumulo database (accumulo.apache.org) is the highest performance open source relaxed consistency database currently available and is widely used for government applications [Byun 2012]. Accumulo is based on the Google Big Table architecture and formally sits on top of the Hadoop distribute file system. Accumulo was developed by the National Security Agency and was released to the open source community in 2011.

Accumulo is designed to handle unstructured data of the type found in document analysis, health records, bioinformatics, social media, computer networks, and computer logs. Often this data is represented as large graphs of nodes and edges. The Graph500 benchmark (Graph500.org)[Bader 2006] is designed to test a computers' ability to process graph data. Graph500 contains a high performance, scalable graph generator that efficiently generates large "power-law" graphs (i.e., graphs with a few nodes with many edges and many nodes with a few edges).

Achieving the full performance of Accumulo requires exploiting its ability to run on parallel computers. This includes insuring that there is sufficient parallelism in the application, load balancing the application across different parts of the system, and minimizing communication between processors. The techniques for achieving high performance on Accumulo are similar to achieving high performance on other parallel computing applications.

The Dynamic Distributed Dimensional Data Model (D4M.mit.edu) [Kepner 2012] provides a uniform framework based on the mathematics of associative arrays [Kepner 2013a] that encompasses both traditional (i.e., SQL) and non-traditional databases. For non-traditional databases D4M naturally leads to a general purpose Accumulo schema that can be used to fully index and rapidly query every unique string in a dataset. The D4M Schema is used across the Accumulo community [Kepner 2013b].

D4M also works seamlessly with the pMatlab (http://www.ll.mit.edu/pMatlab) [Bliss 2006, Kepner 2009] parallel computing environment that allows high performance parallel applications to be constructed with just a few lines of code. pMatlab uses a single-program-multiple-data (SPMD) parallel programming model and sits on top of a message passing interface (MPI) communication layer. SPMD and MPI are the primary tools used in much of the parallel computing world to achieve the highest levels of performance on the world's largest systems (see hpcchallenge.org). These tools can also be used for achieving high performance on the Accumulo database.

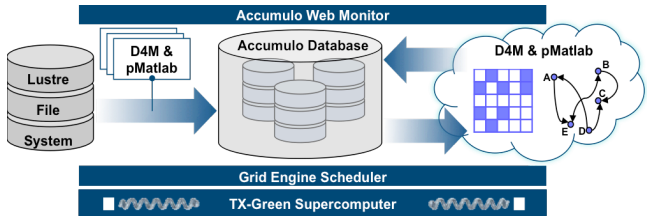

Figure 1. MIT SuperCloud architecture consists of seven components. (1) Lustre parallel file system for high performance file I/O, (2) D4M & pMatlab ingest processes, (3) Accumulo parallel database, (4) D4M & pMatlab analytic processes, (5) Accumulo web monitor page, (6) Grid Engine scheduler for allocating processes to hardward, and (7) the TX-Green supercomputer.

The organization of the rest of this paper is as follows. Section II introduces Accumulo, D4M, pMatlab, and the MIT SuperCloud system used to conduct the performance measurements. Section III describes the Graph500 benchmark data and the benchmark implementation. Section IV describes the specific experiments conducted and the optimizations employed to achieve the measured performance. Section V shows the performance results using the Graph500 benchmark data. Section VI summarizes the results.

## II. TECHNOLOGIES

A variety of technologies were used to conduct the performance measurements. Together, these technologies make up the MIT SuperCloud [Reuther 2013] (see Figure 1). The MIT SuperCloud allows big data applications such as Hadoop and Accumulo to run on a supercomputer system.

### A. Accumulo Database

Accumulo is a key-value store where each entry consists of a seven-tuple. Most of the concepts of Accumulo can be understood by reducing this seven-tuple into a triple consisting of a row, column, and value. Each triple describes a point in a table. Only the non-empty entries are stored in each row, so the table can have an unlimited number of rows and columns and be extremely sparse, which makes Accumulo well-suited for storing graphs.

Accumulo is designed to run on large clusters of computing hardware where each node in the cluster has its own data storage. Accumulo uses the Hadoop Distributed File System (HDFS) to organize the storage on the nodes into a single, large, redundant file system. A table in Accumulo is broken up into tablets where each tablet contains a continuous block of rows. The row values marking the boundaries between tablets are called *splits*. A table can be broken up into many tablets, and these tablets are then stored in HDFS across the cluster. Good performance is achieved when the data and the operations are spread evenly across the cluster. The selection of good splits is key to achieving this goal.

The various Accumulo processes are managed by Zookeeper (zookeeper.apache.org), which is a centralized service for maintaining configuration and naming information, along with providing distributed synchronization and group services.

### B. D4M analytics library

D4M is open source software that provides a convenient mathematical representation of the kinds of data that are routinely stored in spreadsheets and large triple store database. Associations between multidimensional entities (tuples) using string keys and string values can be stored in data structures called associative arrays. For example, in two dimensions, a D4M associative array entry might be

```
        A('alice ', 'bob ') = 'cited '
or      A('alice ', 'bob ') = 47.0
```

The above tuples have a 1-to-1 correspondence with their triple store representations

```
        ('alice ','bob ','cited ')
or      ('alice ','bob ',47.0)
```

Associative arrays can represent complex relationships in either a sparse matrix or a graph form (see Figure 2). Thus, associative arrays are a natural data structure for performing both matrix and graph algorithms. Such algorithms are the foundation of many complex database operations across a wide range of fields [Kepner 2011].

Constructing complex composable query operations can be expressed using simple array indexing of the associative array keys and values, which themselves return associative arrays:

```
A('alice ',:)            alice row
A('alice bob ',:)        alice and bob rows
A('al* ',:)              rows beginning with al
A('alice : bob ',:)      rows alice to bob
A(1:2,:)                 first two rows
A == 47.0                subarray with values 47.0
```

The composability of associative arrays stems from the ability to define fundamental mathematical operations whose results are also associative arrays. Given two associative arrays `A` and `B`, the results of all the following operations will also be associative arrays:

```
A + B    A - B    A & B    A|B    A*B
```

Measurements using D4M indicate these algorithms can be implemented with a tenfold decrease in coding effort when compared to standard approaches [Kepner 2012].

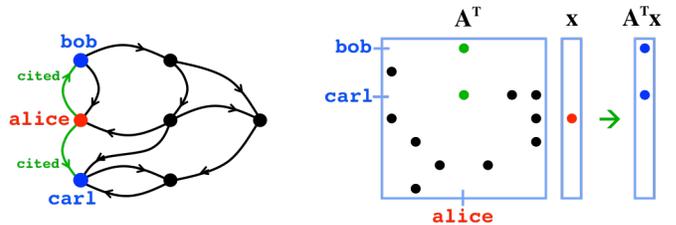

Figure 2. A graph describing the relationship between `alice`, `bob`, and `carl` (left). A sparse associative array **A** captures the same relationships (right). The fundamental operation of graphs is finding neighbors from a vertex (breadth first search). The fundamental operation of linear algebra is matrix vector multiply. D4M associative arrays make these two operations identical. Thus, algorithm developers can simultaneously use both graph theory and linear algebra to exploit complex data.

*C. pMatlab parallel computing library*

pMatlab is open source software that allows a Matlab program(mathworks.com) or a GNU Octave program (octave.org) to be launched in parallel. In a pMatlab program all $N_P$ parallel instances of the program persist for the life of the program, have a unique identifier ($P_{ID}$), and can directly communicate with all the other instances of the programs. The only differences between the instances are the $P_{ID}$s. This parallel programming model is called single-program-multiple-data (SPMD). The communication between each $P_{ID}$ is handled by message passing. In addition, pMatlab provides scalable mechanisms for creating distributed arrays so that each $P_{ID}$ knows exactly which part of the array it owns and where to find all the other pieces.

pMatlab implements the distributed arrays parallel programming model used to achieve high performance on the largest computers in the world. This model gives the application precise control of its computations and communications when running on a parallel computing system.

*D. Lustre parallel file system*

The MIT SuperCloud has two forms of storage: distributed and central. Distributed storage exists on the compute nodes that are used for running Hadoop and Accumulo applications. Central storage is implemented using the open source Lustre parallel file system (lustre.org) on a commercial storage array. Lustre provides high performance data access to all the compute nodes, while maintaining the appearance of a single filesystems to the user. The Lustre filesystem is used in most of the largest supercomputers in the world.

The MIT SuperCloud leverages both types of storage to dynamically start, stop, checkpoint, relocate, and restart (or clone) Accumulo databases by storing their data in the Lustre filesystem when they are stopped. This dynamic database management system allows many more Accumulo databases to be hosted on the system than would otherwise be possible. Groups of users can quickly create their own Accumulo databases to share data amongst themselves without interfering with other groups. In addition, because all the Accumulo instances are running directly on the compute nodes, they can run at maximum performance.

*E. Grid Engine scheduler*

Supercomputers require efficient mechanisms for rapidly identifying available computing resources, allocating those resources to programs, and launching the programs on the allocated resources. The open source Grid Engine software (gridscheduler.sourceforge.net) provides these services and is independent of programming language (C, Fortran, Java, Matlab, …) or parallel programming model (message passing, distributed arrays, threads, map/reduce, …).

The Grid Engine scheduler coordinates the starting and stopping of Accumulo database instances in the MIT SuperCloud. An Accumulo user authenticates using a web page that shows them only the databases they are allowed to access. They can then start and stop any of these databases. When a database is started Grid Engine determines the computing requirements of the database, finds the computing resources, allocates them to the database, copies all the database files to the appropriate computing nodes, assigns dynamic alias domain name entries to the compute nodes, and starts the database processes.

*F. TX-Green hardware*

The TX-Green supercomputer consists of 270 HP servers connected to a single 10 GigE Voltaire core switch. The Lustre central storage system uses a 1 Petabyte DDN and a 0.5 Petabyte Seagate storage array that are directly connected to the core switch. This architecture provides high bandwidth to all the nodes and the central storage. Each server has 32 cores (x86 instruction set), 128 Gigabytes of memory, and 12 Terabytes of storage. The storage is hot-swappable RAID5 so that each node can tolerate one drive failure.

TX-Green is housed in an HP EcoPOD mobile data center that uses ambient air cooling to maximize energy efficiency. The EcoPOD is located near a hydroelectric dam that delivers clean energy that does not contribute green house gases to the environment.

The MIT SuperCloud software stack, which contains all the systems and applications software, resides on every node. Hosting the application software on each node accelerates the launch of large applications (such as Accumulo) and minimizes their dependency on the central storage.

### III. BENCHMARK DESIGN

Accumulo's ability to handle sparse tables makes it well suited for graph applications. Our approach to measuring Accumulo performance begins with generating large graphs, breaking up the graph so it will work well with Accumulo's table structure, and finally creating a parallel program to insert the graph as quickly as possible. High performance insertion of graph data is the first step of many graph applications and is often a key bottleneck. Testing the graph insert performance over a range of system parameters (e.g., number of server processors and number of ingest processes), establishes the upper performance bound on this graph processing step. Other important performance metrics include the query performance and query latency of large graphs. These metrics have been well explored in other work [Sen 2013, Sawyer 2013] and are not explored here.

*A. Graph500 benchmark*

Measuring graph performance begins with generating a graph. The Graph500 benchmark is designed to operate on large graphs. Graph500 has a scalable data generator that can efficiently generate power-law graphs. The number of vertices and edges in the graph are set using a positive integer called the *SCALE* parameter. The number of vertices, *N*, and the number of edges, *M*, are then computed as follows:

$$N = 2^{SCALE} \qquad M = 8\,N$$

For example, if *SCALE* = 17, then *N* = 131072 and *M* = 1048576. The Graph500 generator uses a recursive matrix algorithm [Chakrabarti 2004] to generate a set of starting vertices and ending vertices corresponding to edges in a graph. This graph can then be represented as a large *N*x*N* sparse matrix **A**, where **A**(*i,j*) = 1 indicates an edge from vertex *i* to vertex *j*. Figure 3 shows such a matrix for a *SCALE*=17 graph.

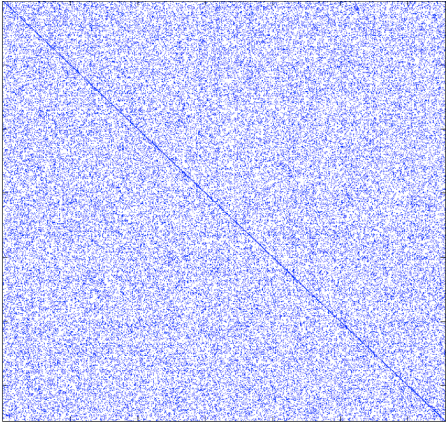

Figure 3. Sparse adjacency matrix representation of *SCALE*=17 Graph500 with ~10% of the edges displayed. The vertices are randomized, so the underlying power-law structure is not visible. The diagonal line shows self-edges that are kept in some applications and filtered out in other applications.

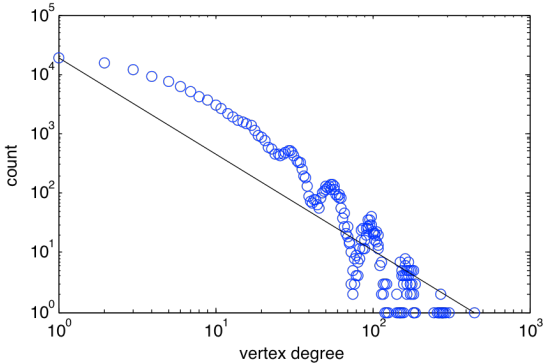

Figure 4. Vertex degree distribution of the graph shown in Figure 3 showing is approximate power-law structure. In this graph there are 18884 vertices with one edge (i.e., a vertex degree of 1) and one vertex with 447 edges. The approximate power law slope of the distribution is: count(vertex degree) $\propto$ (vertex degree)$^{-0.62}$.

A SCALE=17 Graph500 graph is used as a building block for the performance measurements. Three additional parameters determine the overall table that will be ingested. $N_{server}$ is the number of servers used by the Accumulo database and is set to 1, 2, 4, 8, 16, 32, 64, 128, or 216. $N_{ingest}$ is the number of ingest processes per server and is set to 1, 2, 3, 4, 5, 6, or 7. $N_{tablet}$ is the number of tablets per ingest process and is set to a value of 32. Using these parameters, the overall table is constructed by stacking $N_{server}$ $N_{ingest}$ $N_{tablet}$ base graphs to create a single large $N_{row}$ x $N$ table, where

$$N_{row} = N_{server} \; N_{ingest} \; N_{tablet} \; N$$

For the above values, this results in a tables ranging from 4M x 131K with 33M entries to 5.4B x 131K with 43B entries. This approach allows the table to grow with the size of the system and number of ingest processes. In parallel computing, this is referred to as scaled problem (i.e., the problem size grows with computing resources). If the system performance scales linearly, then the ingest time will be constant for any value of $N_{server}$ and $N_{ingest}$.

### B. Domain decomposition

Achieving high performance on any parallel computing problem requires that the data and operations are evenly split amongst the resources. Scaling the table with the number of computing resources simplifies the mapping of specific rows to specific servers, ingest processes, and tablets. Each ingest process generates a base *SCALE*=17 Graph500 table, converts the triple values to strings, and constructs a D4M associative array out of these triples. Next the ingest process computes where its $N_{tablet}$ starting rows should be for this table and assigns new row values to the associative array. Finally, the associative array is inserted into the Accumulo table. This graph construction process results in a perfectly balanced table that is ideal for determining the upper limit on Accumulo ingest performance for graph data.

### C. Parallel program

Implementing the above parallel program requires launching a precise number of ingest processes on each Accumulo server. In addition, each ingest process must know how many total processes are running ($N_P = N_{server} \; N_{ingest}$) and the numerical rank ($P_{ID}$) of its process. These requirements are exactly those of the SPMD programming model. pMatlab conforms to this model. Writing the program using pMatlab makes it easy to implement the above scalable data ingest program. The entire D4M+pMatlab parallel benchmark program consists of four files with 135 total lines of code.

## IV. BENCHMARK IMPLEMENTATION

The implementation of the benchmark required additional performance optimizations to achieve the ultimate performance goal of 100,000,000 database inserts per second. These optimizations fell into two categories: Accumulo optimizations and D4M/pMatlab optimizations. Collecting these optimizations together produced a recipe for running the benchmark.

### A. Accumulo optimization

The Accumulo software used in these performance measurements was Accumulo 1.5, Hadoop 1.1.2, and Zookeeper 3.4.5. In some cases, the optimizations were done to overcome issues that may be fixed in later versions of the software.

By default, Hadoop will replicate all data in its file systems three times which provides a level of redundancy that allows any particular server to fail. In the MIT SuperCloud, the dominant use case is single node Accumulo instances for developers. These instances are unable to utilize this redundancy. In addition, each server uses RAID5 storage that protects from a single disk failure with a lower storage penalty (to store parity information). Throughout these performance measurements, Hadoop was run with no replication to maximize the ingest performance.

At the recommendation of the Accumulo developers, Zookeeper was run with 3 standard instances and 10 follower instances.

Another fault tolerance feature of Accumulo is its write-ahead log that writes all operations to a log prior to performing an insert. The log allows Accumulo to redo an insert if it aborts prior to writing the insert from memory to disk. The MIT SuperCloud allows the Accumulo database to be checkpointed and thus the need for the write-ahead log can be minimized. Accumulo was run with the write-ahead log turned off throughout these performance measurements which increased performance by ~30%.

Accumulo will automatically split and load balance tables as they grow in size. Pre-splitting spreads the table across all the servers at creation. Pre-splitting is required to achieve good ingest performance with an empty table (otherwise it will only exist on one server). Pre-splitting was used throughout these performance measurements. One issue that was encountered is that after creating the pre-splits, they all started out on one server. Accumulo load balanced the splits across its servers at rate of ~50 splits/second, which is more than adequate for normal operation, but can take ~20 minutes for 50,000 pre-splits.

The process by which Accumulo takes entries from memory and initially writes them to disk is called *minor compaction*. By default, Accumulo will limit the maximum number of simultaneous minor compactions on a server to 4. For these performance benchmarks, this value was set to 5, which provided slightly better results.

### B. D4M/pMatlab optimization

To insure that network bandwidth was not limiting Accumulo ingest performance, the program was run so that each ingest process was inserting data into tablets that were local to the Accumulo server it was running on. This was accomplished by querying Accumulo for the server locations of each split and saving these locations to disk. Each ingest process then read in this file and discovered the tablets that were local to it. Furthermore, if multiple ingest processes were on a server, they used their $P_{ID}$ and their hostname to evenly divide up the splits on the same server.

D4M can insert an entire associative array into a table with a simple "put" command. Inside the D4M put command, the associative array is divided into blocks that are individually inserted using the Accumulo batch writer API. The block size was chosen to be 500 kilobytes, which typically provides optimal ingest performance.

### C. Benchmarking recipe

The steps for running each benchmark were as follows and involves a setup phase and an execution phase.

Setup Phase

(1) Start Accumulo on $N_{server}$ servers.

(2) Set max compactions to five via
    `tserver.compaction.minor.concurrent.max=5`.

(3) Create the table.

(4) Disable write ahead log via `table.walog.enabled=false`.

(5) Create all the table splits and wait for them to load balance across all the servers.

(6) Retrieve all the splits and their corresponding servers and write to a file in the central filesystem.

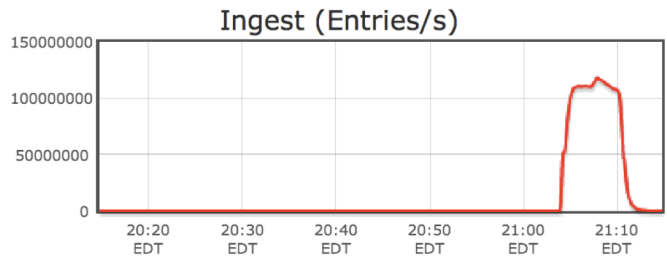

Figure 5. Accmulo ingest performance vs time as recorded by the Accumulo web monitor page. The benchmark runs for ~300 seconds. The rise and fall in the curve is mostly due to the Accumulo averaging window.

(7) Launch $N_{ingest}$ processes on each Accumulo server using pMatlab.

Execution Phase (Each ingest process executes the following.)

(1) Read in split file from the central filesystem.

(2) Use $P_{ID}$, $N_P$ and hostnmae to find splits that are local to server and divide these splits evenly among all ingest processes.

(3) Generate *SCALE*=17 Graph500 graph and inserts into a D4M associative array.

(4) Use each split point value to offset the rows of the associative array.

(5) Insert the associative array into the Accumulo table.

## V. PEFORMANCE RESULTS

The benchmark allows high performance ingest with minimal ramp up time (see Figure 5). After the setup phase, a typical run time for the program was 300 seconds. The benchmark program was run over a range of Accumulo database instantiations across the following number of servers: 1, 2, 4, 8, 16, 32, 64, 128, 216; and different numbers of ingest processes per server: 1, 2, 3, 4, 5, 6, 7. Figure 6 shows the ingest performance vs. ingest processes for all the different Accumulo configurations. Linear performance was achieved across all the dimensions culminating in a peak performance of 115,000,000 entries/second on a 216 node Accumulo database with 1,296 ingest processes. On average, 100,000 entries/second was achieved per ingest process across the range of measurements. Likewise, a maximum performance of 500,000 entries/second per server node was also achieved. For comparison, Figure 7 shows the maximum ingest rate as function of the total number of nodes used for Accumulo, Cassandra, and Oracle. Different benchmarks and hardware platforms were used for these results, so specific comparisons must be made with care. Overall, it appears that Accumulo provides the highest ingest performance of all of these technologies.

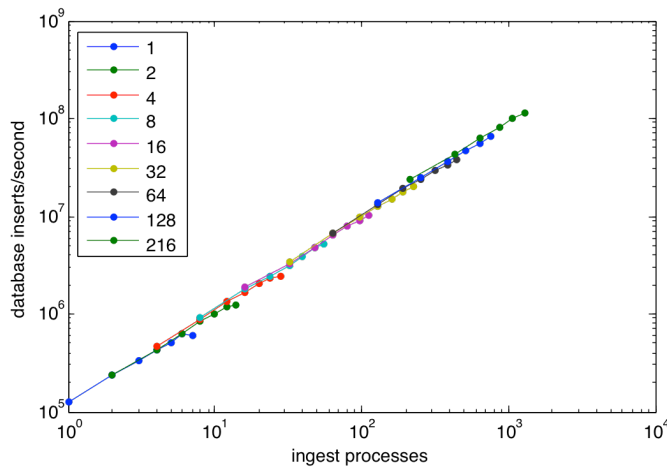

Figure 6. Ingest performance vs. number of ingest processors for different Accumulo databases with different numbers of servers (see legend) demonstrating linear performance scaling.

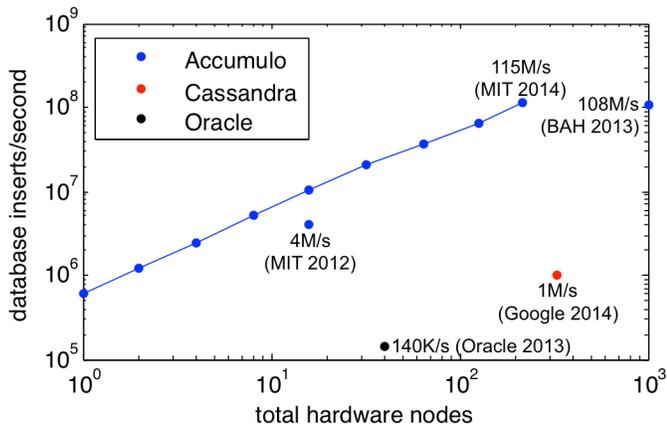

Figure 7. Maximum ingest performance of various database technologies on *different* benchmarks: (MIT 2014) Graph500 data [this paper], (MIT 2012) computer network data [Byun 2012], (BAH 2013) Accumulo continuous test suite [Sen 2013], (Google 2014) random data [Filho 2014], and (Oracle 2013) TPC-C benchmark [TPC 2013].

## VI. SUMMARY

The Apache Accumulo database is an open source relaxed consistency database that is widely used for government applications. Accumulo is designed to deliver high performance on unstructured data such as graphs of network data. This paper measured the performance of Accumulo using data from the Graph500 benchmark. D4M and pMatlab software were used to implement the benchmark on a 216 node cluster running the MIT SuperCloud software stack. A peak performance of over 115,000,000 database inserts per second was achieved, which is 100x larger than the highest previously published value for any other database. The performance scales linearly with number of ingest clients, number of database servers, and data size. This performance was achieved by adapting several supercomputing techniques to this application: distributed arrays, domain decomposition, adaptive load balancing, and single-program-multiple-data programming.